\documentclass[aip,reprint]{revtex4-1}
\usepackage{graphicx}
\usepackage{dcolumn}
\usepackage{bm}
\usepackage{amsmath}
\usepackage{mathrsfs}
\usepackage{empheq}
\usepackage{url}
\usepackage{natbib}
\usepackage{aas_macro}
\usepackage{bigints}
\usepackage{braket}
\usepackage{hyperref}
\usepackage{subfigure}
\usepackage[applemac]{inputenc}
\usepackage[usenames, dvipsnames]{color}

\begin{document}

\preprint{AIP/123-QED}

\title{Energy transfer and electron energization in collisionless magnetic reconnection for different guide-field intensities.}

\author{F. Pucci}
\email{fpucci@nifs.ac.jp, fpucci@princeton.edu}
\affiliation{National Institute for Fusion Science, National Institutes of Natural Sciences,
Toki 509-5292, Japan}
\affiliation{Princeton Plasma Physics Laboratory, Princeton University, Princeton, 08543-0451,New Jersey, USA}
\author{S.Usami}%
\affiliation{National Institute for Fusion Science,Toki 509-5292, Japan}

\author{H. Ji}%
\affiliation{Princeton Plasma Physics Laboratory, Princeton University, Princeton, 08543-0451,New Jersey, USA}
\affiliation{ Department of Astrophysical Sciences, Princeton University, Princeton, New Jersey 08544, USA}%
\author{X. Guo}%
\affiliation{ The University of Tokyo, 5-1-5 Kashiwanoha, Kashiwa, Chiba 277-8561, Japan
}%

\author{R. Horiuchi}%
\affiliation{National Institute for Fusion Science,Toki 509-5292, Japan}

\author{S. Okamura}%
\affiliation{National Institute for Fusion Science, National Institutes of Natural Sciences,
Toki 509-5292, Japan}

\author{W. Fox}%
\affiliation{Princeton Plasma Physics Laboratory, Princeton University, Princeton, 08543-0451,New Jersey, USA}

\author{J. Jara-Almonte}%
\affiliation{Princeton Plasma Physics Laboratory, Princeton University, Princeton, 08543-0451,New Jersey, USA}

\author{M. Yamada}%
\affiliation{Princeton Plasma Physics Laboratory, Princeton University, Princeton, 08543-0451,New Jersey, USA}

\author{J. Yoo}%
\affiliation{Princeton Plasma Physics Laboratory, Princeton University, Princeton, 08543-0451,New Jersey, USA}

\date{\today}

%
%
%
\begin{abstract}

Electron dynamics and energization are one of the key components of magnetic field dissipation in collisionless reconnection. In 2D numerical simulations of magnetic reconnection, the main mechanism that limits the current density and provides an effective dissipation is most probably the electron pressure tensor term, that has been shown to break the frozen-in condition at the x-point. In addition, the electron-meandering-orbit scale controls the width of the electron dissipation region, where the electron temperature has been observed to increase both in recent Magnetospheric Multiple-Scale (MMS) observations as well as in laboratory experiments, such as the Magnetic Reconnection Experiment (MRX).  By means of two-dimensional full-particle simulations in an open system, 
 we investigate how the energy conversion and particle energization depend on the guide field intensity. We study the energy transfer from magnetic field to the plasma in the vicinity of the x-point and close downstream regions, ${\bf E}\cdot {\bf J}$ and the threshold guide field separating two regimes where either the parallel component, $E_{||}J_{||}$, or the perpendicular component, ${\bf E}_{\perp}\cdot {\bf J}_{\perp}$, dominate the energy transfer, confirming recent MRX results and also consistent with MMS observations. We calculate the energy partition between fields, kinetic, and thermal energy of different species, from electron to ion scales, showing there is no significant variation for different guide field configurations. Finally we study possible mechanisms for electron perpendicular heating by examining electron distribution functions and self-consistently evolved particle orbits in high guide field configurations.

\end{abstract}

\pacs{Valid PACS appear here}
\maketitle

%
%
\section{Introduction}\label{sec:0} 

Magnetic reconnection is thought to play a key role in explosive phenomena in space and laboratory plasmas, such as solar flares, substorms in the Earth's magnetosphere, and disruptions in laboratory fusion experiments. In all these events, energy stored in the magnetic field is released on fast time-scales principally into thermal and non-thermal energies of the ambient particles. 
A kinetic study of reconnection dynamics is required in most high-temperature natural and laboratory plasmas, since in such collisionless systems dissipation occurs at particle gyration scales. The nonlinear evolution of a current sheet may involve single or multiple x-points (in the case of multiple plasmoid formation) \cite{Pei_etal:2001,Daughton:2016,Drake:2003_Sci, Horiuchi_etal:2018}, in which case the island dynamics affects particle acceleration. After an initial energization at the reconnection x-point, particles enter cavities, interact with islands and are reflected and scattered by adiabatic mechanisms, like Fermi acceleration, as well as non adiabatic processes. Nonetheless the initial acceleration occurs at the x-point serving as an injection mechanism, so it is fundamental to investigate how energization occurs in this region and in the nearby outflow region. 
A detailed study of the energy transfer from field to particles in antiparallel reconnection was carried out by \citet{YamadaNature2014, Yamada2015} with the Magnetic Reconnection Experiment (MRX), showing that the energy deposition rate on electrons, calculated as ${\bf J_e} \cdot {\bf E}$ where $J_e$ is the electron current, occurs in a region surrounding the x-point wider than predicted by 2D numerical simulations, so that a notable rise of electron temperature (up to $50\%$) is measured over an area that is much wider than the electron diffusion region. Recently magnetospheric observations from the Magnetospheric Multi-Scale mission (MMS) have identified guide field reconnection events, where the symmetric quadrupolar structure of the magnetic field is altered, and the associated reconnection electric field and temperature is measured. \citet{Eriksson_etal:2016} reported MMS observations of a large guide field magnetic reconnection event where the guide field amplitude being approximately 4 times the reconnecting field. One of MMS satellites (MMS3) detected a significant parallel electric field throughout the electron diffusion region (EDR) with significant parallel heating. \citet{Wilder_etal:2017} also observed a peak in the electron temperature during the crossing of the EDR by an MMS satellites with guide field $B_z \sim B_0$, i.e. comparable to the reconnecting component. \citet{Genestreti_et_al:2017}  compared guide field reconnection configurations to determine how the rate of work done by the electric field varies with shear angle.
Particle heating was shown to be sensitive to guide field variation in laboratory plasmas. \citet{Ono_etal:2012} in their merging spheromak experiment, observed that electrons are heated inside the current sheet, as evidenced by the measured electron temperatures and plasma flow. They also noted that the extent of ion heating depends on the guide field magnitude. \citet{Usami_et_al:2018} by means of particle simulations, found that the ion temperature rises mainly in the downstream, where ring-like structures of ion velocity distributions are formed. The ion temperature profiles in the high guide field PIC simulations are in a qualitative agreement with the TS-3 experiment \citet{Ono_etal:2012}.
\citet{Tanabe_etal:2015} found that an increment in the toroidal guide field results in a more peaked electron temperature profile at the x-point (a similar trend is found in the MRX experiment \cite{Fox_et_al:2017}), while the ion temperature profile forms double peaks in the outflow region, where the peaks seem unaffected from guide field changes. \citet{Drake_et_Swisdak:2014} simulated a strong guide field case, finding out that the dominant heating of thermal heavy ions during guide field reconnection results from pickup behavior of heavy ions during their entry into reconnection exhausts and dominantly produces heating perpendicular rather than parallel to the local magnetic field and in general parallel heating in the guide field case is strongly reduced with respect to the anti parallel case.
In addition, recent measurements from the MRX experiment, in agreement with MMS observations, show that higher guide fields lead to a higher contribution of parallel energy transfer $E_{||} J_{||}$, with respect to perpendicular energy transfer ${\bf E_{\perp}}\cdot {\bf J_{\perp}}$,  to the total energy transfer ${\bf E}\cdot {\bf J}$ (\citet{Fox_et_al:2018}).\\
Numerical studies of guide field reconnection have been carried out in 2D as well as in 3D (e.g.\citet{Lapenta_et_al:2014}), also in asymmetric configurations \citet{Pritchett_Mozer:2009}. A significant range of guide field variations $B_{z}/B_0 \sim 0-4 $ is needed for comparison with recent MMS observations.
In Sect. II of this paper we describe the simulation setup, followed by the descriptions of the results on how the energy transfer from magnetic field to particles, in the vicinity of the x-point, changes for different values of the guide field in Sect. III. We compare the results with the MRX measurements and also MMS observations for antiparallel and guide field configurations. In Sect. IV, we discuss different energization regions and energy redistribution between ions, electrons and electromagnetic field. In Sect. V, we discuss energization mechanisms for electrons and provide a statistical study of self consistently evolved particles to analyze electron temperature for different guide field configurations.

%
%
\section{Simulation setup.}\label{sec:I} 

\begin{figure}
\includegraphics[width=80mm]{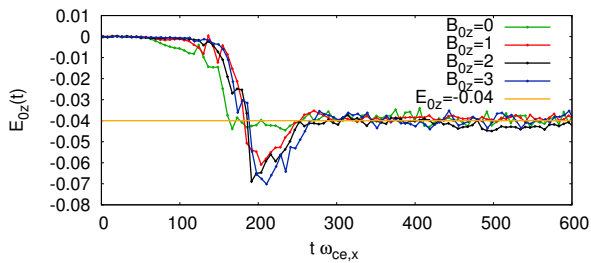}
\caption{Reconnection electric field (orthogonal to the reconnection plane) at the x-point as a function of time, for different guide field values, runs $G0-G3$.}
\label{Fig:ref_field}
\end{figure}
\begin{figure}
\includegraphics[width=80mm]{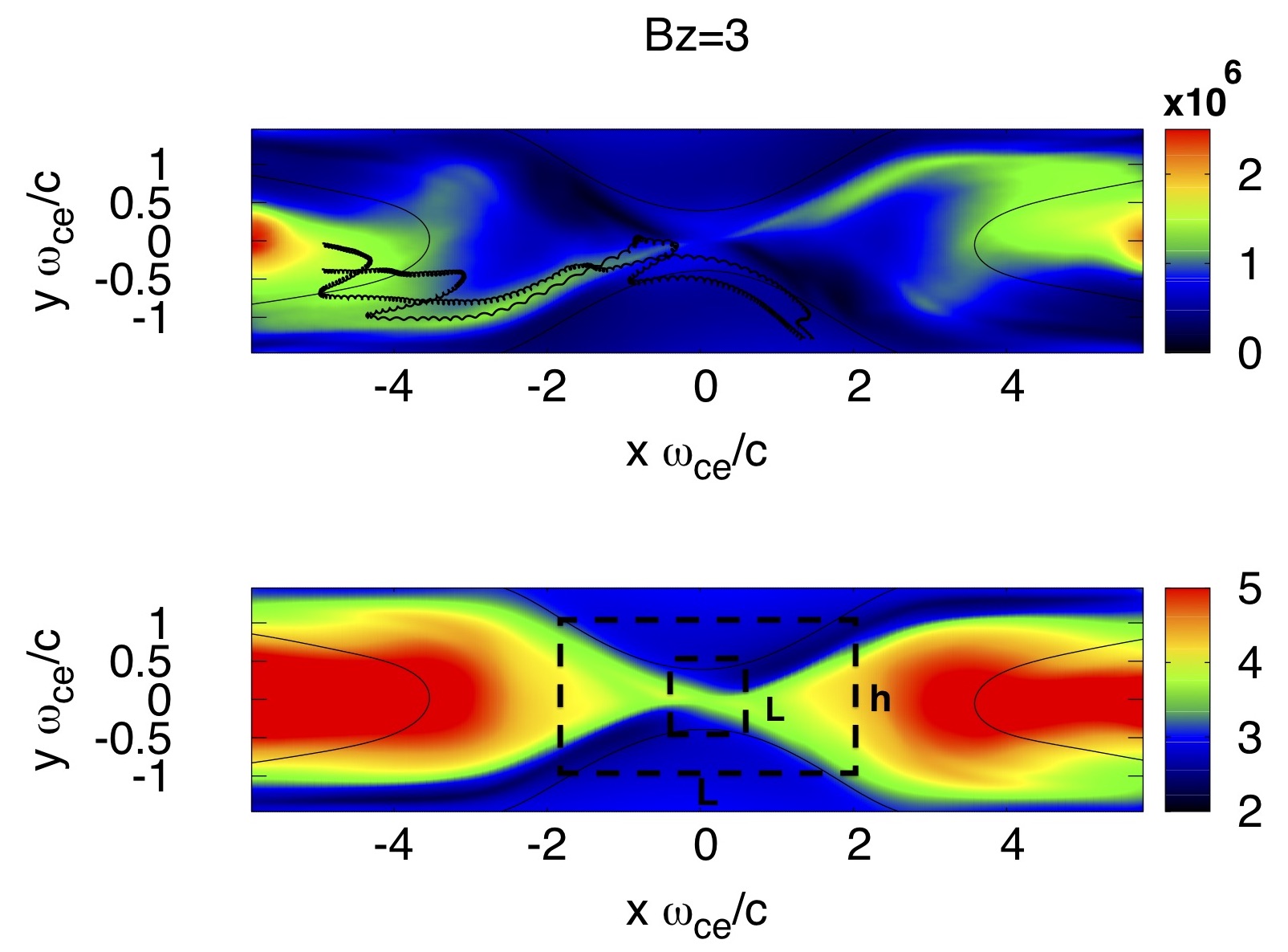}
\caption{2D profiles for $B_{0z}=3$ of: (top) electron number density with two typical electron orbits superimposed (bottom) out of the plane component of the magnetic field $B_{z}$. The classical quadrupolar structure is altered by the presence of a strong guide field.}
\label{Fig:orbite00}
\end{figure}

We carry out two dimensional particle-in-cell simulations of driven magnetic reconnection using the PASMO code \cite{Horiuchi_Sato:1997, Pei_etal:2001, Ishizawa_Horiuchi_2005, Horiuchi_etal:2014}. The system is subject to an external driving flow, obtained by imposing an electric field at the two upstream boundaries (y = $\pm y_b$), perpendicular to the magnetic field, which pushes particles into the simulation domain via the ${\bf E}\times{\bf B}$ drift. The driving electric field is described in \citet{Pei_etal:2001}, see in particular Fig. 1 of the latter paper. In the outflow direction (x-axes) we employ open boundary conditions (BCS) so that we can achieve a steady state by avoiding that the reconnection jets might propagate across the boundaries and back into the simulation domain, affecting the dynamics, as naturally occurs with periodic BCS.
The initial condition consists of an equilibrium that depends only on the y-coordinate with an antiparallel magnetic field along the x-axis and a uniform guide field along the z-axis:
\begin{eqnarray} 
\label{eq:equilibriumMAGFILD}
&&{\bf B}=B_{0} \,\mathrm{tanh}(y/a)\,{\bf e}_{x} + B_{0z}\, {\bf e}_{z}\\[2ex]
&&P=B_{0}^2/(8\pi) \,\mathrm{sech}^2(y/a)+P_{0};
\end{eqnarray}
here, $P$ is the pressure, due to a part $P_{0}$ coming from background particles of density $0.35 n_0$, with $n_0$ the particle density at the neutral sheet, $B_{0z}$ is a constant guide field, while $a$ defines the scale of the gradient of magnetic field.  The isotropic plasma pressure balances the upstream magnetic pressure. We normalize time to $ 1/\omega_{ce}$, velocities to the speed of light $c$, the length-scales to $c/ \omega_{ce}$, magnetic and electric fields to the asymptotic value of the reconnecting field, $B_{0}$. The initial particle distribution is a shifted Maxwellian with a spatially constant temperature ($T_e=T_i$) and an average particle velocity equal to the diamagnetic drift velocity. Quantities are assumed to be uniform in the direction perpendicular to the plane of the equilibrium magnetic field, i.e. $\partial /\partial z=0$.  We have carried out a series of runs under various guide-field conditions to quantify the energy deposition regions and mechanism, with a mass ratio of $m_i/m_e=100$ and $T_i/T_e=1$. The domain size is $[11.73 \times 2.93] \, c/\omega_{ce}$ and the ratio $\omega_{pe}/\omega_{ce}=9$. In Tab. \ref{tabellaPIC1}, we summarize the main parameters. 
%
%
%
\begin{table}[h]
\centering
\begin{tabular} {|c|c|c|c|c|c|c|c|c|}
\hline
   {\bf Name}              &$N_x$        & $N_y$       & Particles      &$m_i/m_e $ &$\omega_{pe}/\omega_{ce}$ &$E_{0z} $ &$B_{0z}$&$a$\\ 
\hline
G0    & $768$        &$385$              &$14336 \times 10^{3} $       &$100$ &$9$ &$0.04$ &$0$ &$0.355$\\
G02  & $768$        &$385$               &$14336 \times 10^{3} $  &$100$ &$9$ &$0.04$ &$0.2$ &$0.355$\\
G05    & $768$        &$385$              &$14336 \times 10^{3} $    &$100$ &$9$ &$0.04$ &$0.5$ &$0.355$\\
G08    & $768$        &$385$              &$14336 \times 10^{3} $  &$100$ &$9$ &$0.04$ &$0.8$ &$0.355$\\
G1  & $768$        &$385$               &$14336 \times 10^{3} $  &$100$ &$9$ &$0.04$ &$1$ &$0.355$\\
G2    & $768$        &$385$              &$14336 \times 10^{3} $    &$100$ &$9$ &$0.04$ &$2$ &$0.355$\\
G3    & $768$        &$385$              &$14336 \times 10^{3} $  &$100$ &$9$ &$0.04$ &$3$ &$0.355$\\
 \hline
\end{tabular}
\caption{\textit{Simulation parameters. $N_x$ and $N_y$ are the grid sizes on the x and y axes respectively, the number of active particles for each species, mass ratio $m_i/m_e$, ratio between the electron plasma frequency and cyclotron frequency, asymptotic value of the driving electric field, guide field and initial thickness of the current sheet.}} 
\label{tabellaPIC1}
\end{table}
%
%
%
%
In Fig.\ref{Fig:ref_field} we plot the reconnection electric field component at the x-point (orthogonal to the plane where magnetic reconnection occurs) as a function of time, for different value of the guide field (simulations $G0-G3$). In an initial transient phase $(t < 300)$  the electric field reaches a (negative) minimum value whose absolute value increases as the guide field increases. This is due to magnetic flux accumulation in the x-point region, resulting in an initial enhancement of the reconnection rate. After the initial transient phase $(t > 300)$ the system reaches a stationary state. \citet{Horiuchi_Sato:1997} found that under the influence of an external driving flow, the electron current layer thickness decreases with the guide field, and the reconnection rate is determined by the driving electric field. Indeed, in the stationary state the reconnection electric field levels out asymptotically to  $E_{z}= -0.04$.
In Fig.\ref{Fig:orbite00} we show the 2D profile of the electron number density and the Hall quadrupolar magnetic field structure (in the z-direction orthogonal to the reconnection plane) over the full simulation domain, for the case of guide field $B_{0z}=3$.  The separatrices present a strong asymmetric structure of high and low densities. Similar structure has been observed in MRX by \citet{Fox_et_al:2017}. We also note that the symmetric quadrupolar structure is altered by the presence of the out-of-plane guide field \cite{Horiuchi_Sato:1997, Pritchett:2001b}. Superimposed over the electron number density in Fig. \ref{Fig:orbite00}  two typical electron orbits are shown that will be discussed in Sec. V.
%
%
%
%
%
\section{Energy conversion from fields to particles.}\label{sec:II} 
In this section we will discuss the energy transfer from the fields to plasma at the different guide fields, first using a single fluid approach, calculating ${\bf E}\cdot {\bf J}$ and related quantities. We then quantify the transfer of electromagnetic energy to the plasma in the two fluid framework, the latter being the proper general approach in a kinetic study, as we are particularly interested in electron energization.\\
%
\subsection{The energy deposition in the laboratory frame ${\bf E}\cdot {\bf J}$.}\label{sec:II} 

\begin{figure*}
\includegraphics[width=130mm]{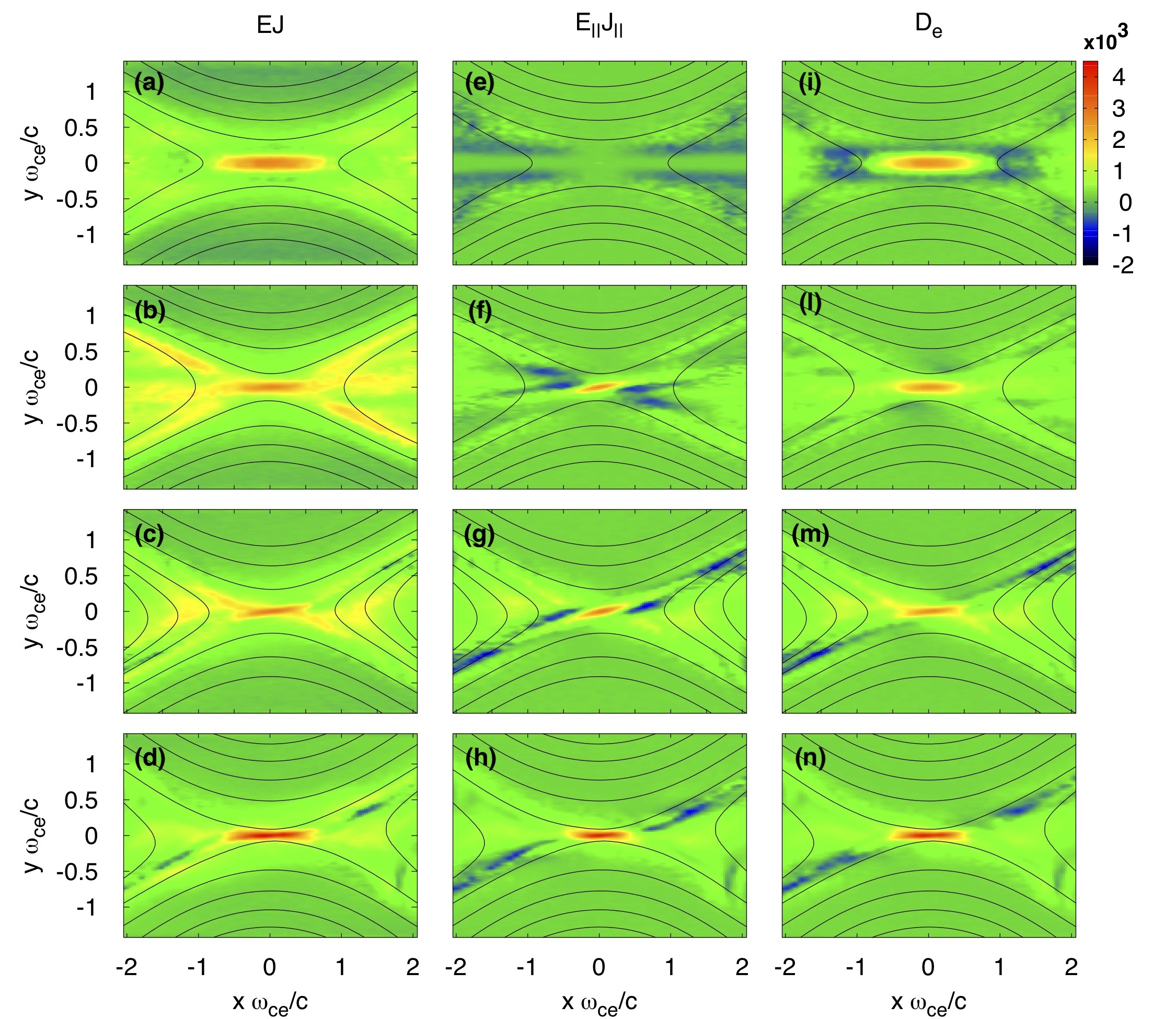}
\caption{2D profiles for simulations $G0-G3$ (increasing guide field from top to bottom) of: (a-d) the energy transfer ${\bf E}\cdot {\bf J}$, (e-h) $E_{||}J_{||}$, (i-n) $D_e$ dissipation in the electron reference frame.}
\label{Fig:dissipation1}
\end{figure*}

\begin{figure}
\includegraphics[width=70mm]{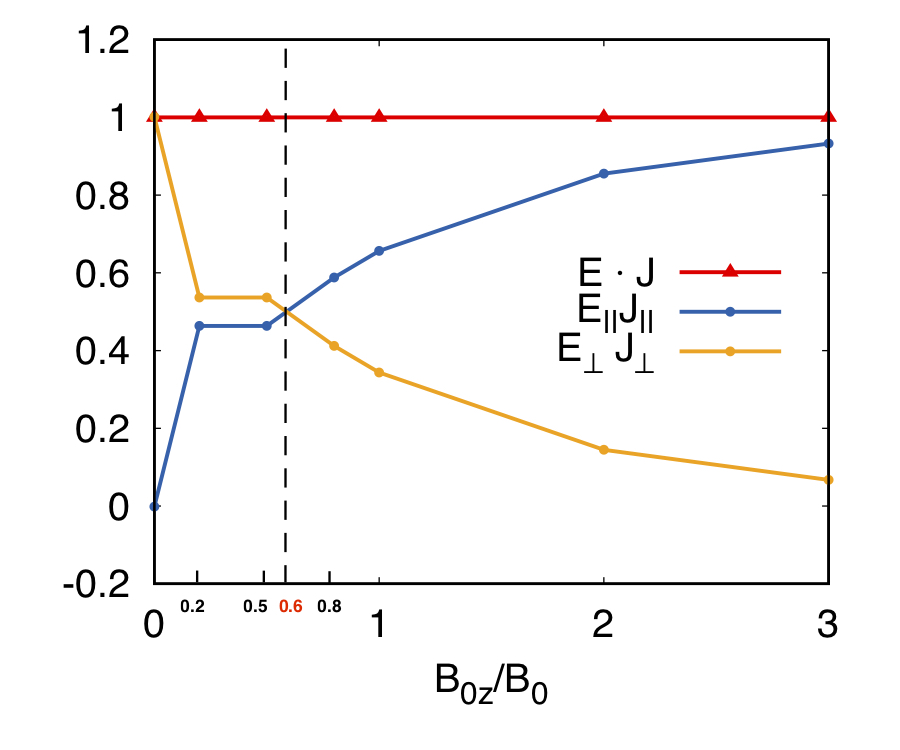}
\caption{Magnetic energy transfer to the plasma ${\bf E}\cdot {\bf J}$ (in logarithmic scale) for different values of the guide field, simulations $G0-G3$ (horizontal axis), compared with parallel $E_{||}J_{||}$ and perpendicular contribution ${\bf E}_{\perp}{\bf J}_{\perp}$, normalized to ${\bf E}\cdot {\bf J}$.}
\label{Fig:EJ_thresh}
\end{figure}

\begin{figure}
\includegraphics[width=80mm]{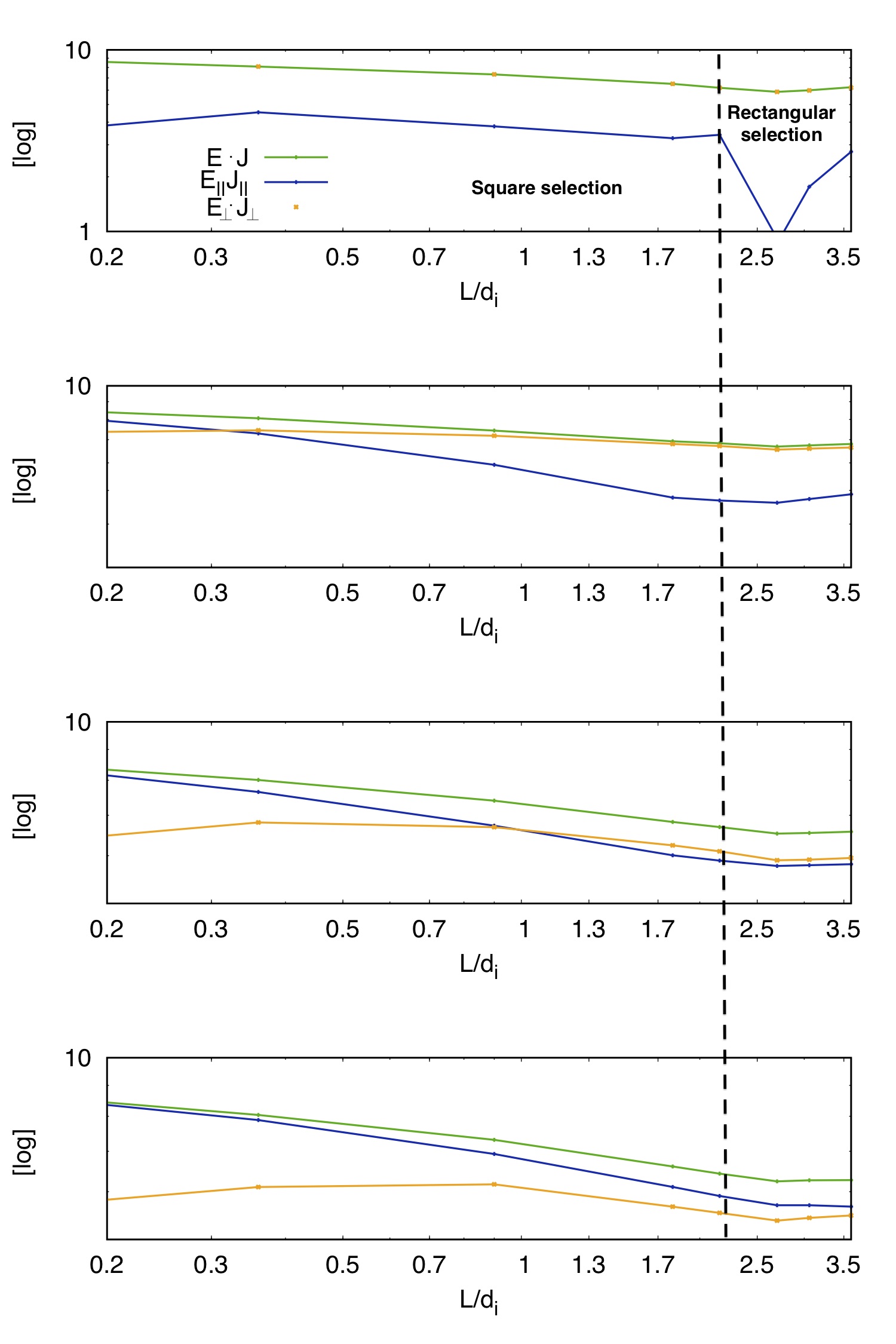}
\caption{Magnetic energy transfer to the plasma (logarithmic scale), ${\bf E}\cdot {\bf J}$ for different values of the guide field (corresponding to simulations $G0-G3$ as indicated in the panels), compared with parallel $E_{||}J_{||}$ (blue solid line) and perpendicular contribution ${\bf E}_{\perp}\cdot{\bf J}_{\perp}$ (orange solid line), integrated over different areas centered on the x-point. For $L < 2d_i$, the areas over which we integrate are squares $[L\times L]$. The linear size $L$ of each square, normalized to the ion skin depth, is indicated on the horizontal axis.
For $L > 2d_i$, the areas are rectangles $[L\times h]$, with fixed height $h = 2d_i$, varying $L$ as indicated on the x axis.}
\label{Fig:EJ_thresh2}
\end{figure}

Fig. \ref{Fig:dissipation1} (left column) shows the energy deposition on the plasma, ${\bf E}\cdot {\bf J}$. A positive value indicates that magnetic energy is converted into particle energy, while for negative values energy goes to the fields \cite{BirnHesse:2005}. 
The energy transfer from the fields to the plasma at the x-point is enhanced as the guide field increases. Recent measurements from the MRX experiment, in agreement with MMS observations, show that higher guide fields lead to a higher contribution of parallel energy transfer $E_{||} J_{||}$, with respect to perpendicular energy transfer ${\bf E_{\perp}}\cdot {\bf J_{\perp}}$,  to the total energy transfer ${\bf E}\cdot {\bf J}$ (\citet{Fox_et_al:2018}). Parallel energy transfer becomes dominant in the MRX experiment already at $B_{z} =0.8$, suggesting a transition from perpendicular to parallel dominated energy transfer between $B_{z}=0$ and $B_{z}=0.8$. Simulations $G0-G3$ show a qualitative agreement with this result, as can be seen from the first two columns of Fig. \ref{Fig:dissipation1}. In the latter we show a zoom of the reconnection region $[L_x\, \times L_y]= [4\, \times 3]c/\omega_{ce}$ centered at the reconnection region, to better compare with \citet{Fox_et_al:2018} data from MRX. In first column we show the total energy transfer ${\bf E}\cdot{\bf J}$ while in the second column we show the parallel contribution ${\bf E}_{||}\cdot{\bf J}_{||}$. Integrating ${\bf E}\cdot {\bf J}$ and the parallel and perpendicular contribution around the x-point, within the electron diffusion region, for simulations $G0-G3$ in Fig. \ref{Fig:EJ_thresh}, we can see that the threshold is confirmed to be at $B_{0z}< 1$ \citet{Pucci_et_al:2017AGU, Li_et_al:2018}. A more detailed analysis with guide field configurations $0<B_{0z}< 1$ indicate the threshold value is around guide field $B_{0z}=0.6$. The transition between the zero guide field configuration and $B_{0z}=0.2$ appears to be sharp. Further investigation of the threshold value are strongly subject to the integration area, as we can see from Fig. \ref{Fig:EJ_thresh2}. Indeed while in \citet{Li_et_al:2018} the integration is over a box $[L_{x} \times L_{y}]= [200 \times 40] d_i$, here we discuss the integration close to the x-point region, on box with linear size $L$ ranging from the electron skin depth to about three ion inertial lengths. In Fig. \ref{Fig:EJ_thresh2} we normalized the length-scales to the ion skin depth $d_i$ so that $L/d_i=0.2$ means $L=2\,d_e$.
We find that for moderate guide fields $B_{0z}=1,2$ the fraction of energy converted by parallel or perpendicular fields depends on the volume of the region near the x-point analyzed. For small volumes ($L < d_i$) it is dominated by parallel energy transfer, while on larger scales the perpendicular energy transfer is dominant. For the zero guide field case we confirm that the energy transfer is perpendicularly dominated independently of the integration area, while for $B_{0z}=3$ the parallel energy transfer is dominant. In addition the current distribution is found to be strongly dependent on the mass ratio. In order to reproduce the actual structure of the dissipation region, real mass ratio simulations should be performed and compared to the results we present here. \citet{Le_et_al:2013} provided an important parametric study for  guide fields in the range $B_{0z}=0-0.8$ and by varying the mass ratios they showed that x-point structure and jets vary significantly with the mass ratio.
Nonetheless we confirm the presence of significative values of $E_{||} J_{||}$ at the x-point is a characteristic of guide field reconnection. The latter result, seen in  Fig. \ref{Fig:EJ_thresh2}, suggests that in observations such as those described by \citet{Phan_et_al:2018} it is areas close to the reconnection region that have been probed, giving an idea of the location of the x-point in observation data. 
%

\subsection{Bulk energy and internal energy: the electron dissipation measure.}\label{sec:3B} 

While ${\bf E}\cdot {\bf J}$ gives the energy transfer from the fields to the plasma, i.e. it is related to either bulk flow or thermal energy variations, if we consider only the internal energy evolution $u=3/2P$ (assuming for sake of simplicity a scalar pressure)
\begin{equation}
\label{eq:tfluid_velocity}
\partial_t u = -\nabla \cdot \left[\left(u +P\right){\bf v}\right] + {\bf v} \cdot \nabla P+  {\bf E}'\cdot{\bf J}\\
\end{equation}
where ${\bf v}={\bf v}_i -{\bf J}\dfrac{m_e}{n_e(m_i+m_e)}$ is the fluid velocity and $\bf E'={\bf E}+{\bf v}\times {\bf B}$ is the electric field calculated in the fluid rest frame. Using the definition for the fluid velocity we get
\begin{eqnarray}
\label{eq:energy_transfer_total_prime}
&&{\bf E}'\cdot{\bf J}={\bf E}\cdot{\bf J} + ({\bf v}_i \times{\bf B})\cdot {\bf J}.
\end{eqnarray}
In a two fluid model (electrons and ions), the energy transfer calculated in the electron frame (a similar calculation can be provided for the ion frame) may be written as  \citet{Zenitani_etal:2011} 
\begin{equation}
\label{eq:Edissipation_total}
{\bf D}_e=\gamma_e \left[{\bf J}\cdot({\bf E}+{\bf v}_e\times{\bf B})-\rho_c({\bf v}_e\cdot {\bf E})\right],
\end{equation}
where in our case $\gamma_e =[1-(v_e/c)^2]^{-1/2} \simeq 1$ is the Lorentz factor and $\rho_c=n_i-n_e$ is the charge separation. We see that, where charge neutrality holds, i.e. where $\rho_c$ is negligible, ${\bf E' \cdot J} = {\bf D}_e$. So under this condition ${\bf D}_e$ is a good indicator for changes in the internal energy of the plasma.  Charge neutrality is approximately valid throughout the simulation domain with the possible exception of the high density separatrix region in the high guide field case \cite{Guo_etal:2017}. We verified that the difference between ${\bf E' \cdot J}$ and ${\bf D}_e$ resulted to be negligible. Fig. \ref{Fig:dissipation1} (right column) shows the parameter $D_e$ in simulations $G0-G3$. As the guide field increases and in particular for the case of strong guide field $B_{0z}=3$ the value at the x-point is twice as the case of zero guide field configuration.
We can see that positive $D_e$ is localized at the x-point, while it assumes negative values in the outflow region immediately close to the x-point in the zero guide field case (as remarked by \citet{Zenitani_etal:2011}). 

\subsection{High guide field configuration: electron acceleration and generation of an electrostatic field.}\label{sec:IIIC} 

\begin{figure}
\includegraphics[width=80mm]{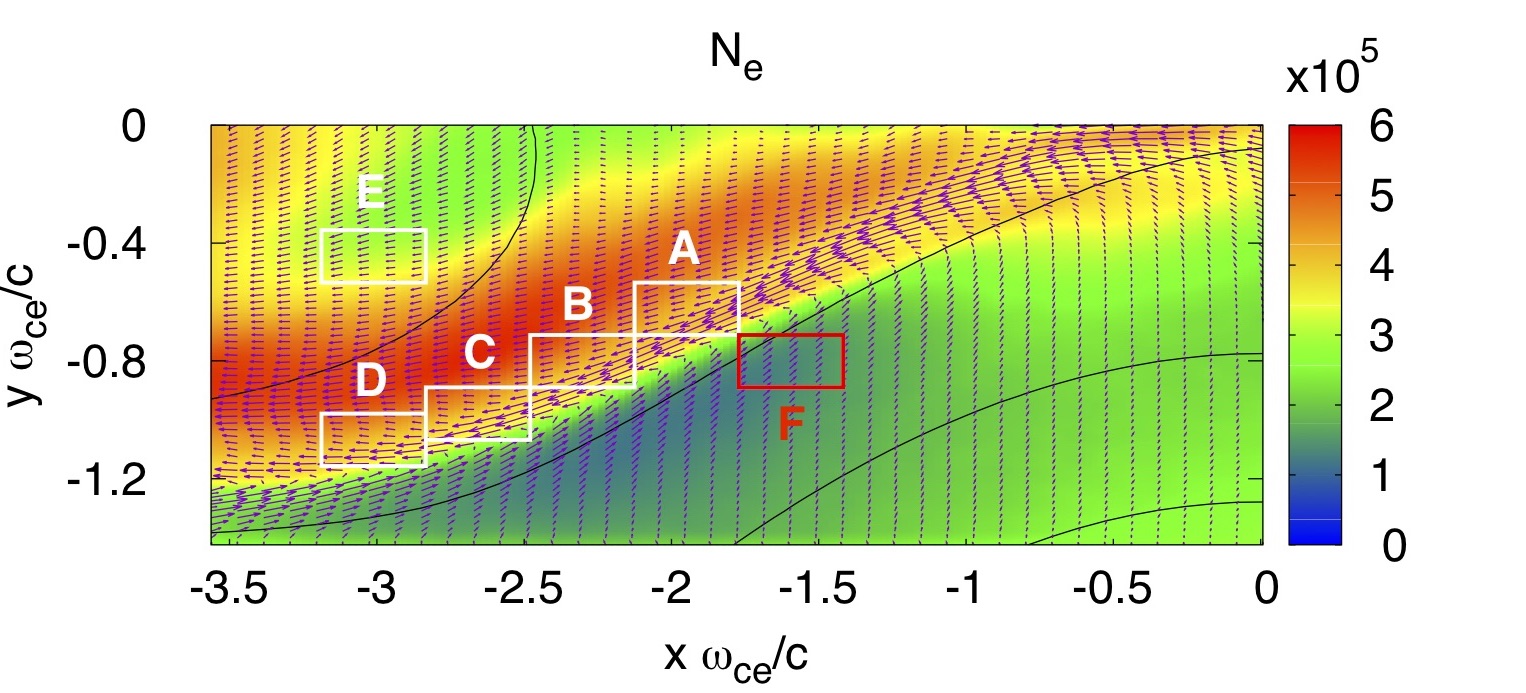}
\caption{Simulation G3: vector plot of the electron fluid velocity components $v_{ex}$ and $v_{ey}$, color coded the electron density $n_e$, averaged between $\sim 6$ electron gyration orbits. Superimposed black lines represent magnetic field lines, while box indicate the region where distribution functions are calculated, see Fig.\ref{Fig:distribution_functions1}-\ref{Fig:distribution_functions2}.}
\label{Fig:boxes}
\end{figure}
\begin{figure*}
\includegraphics[width=130mm]{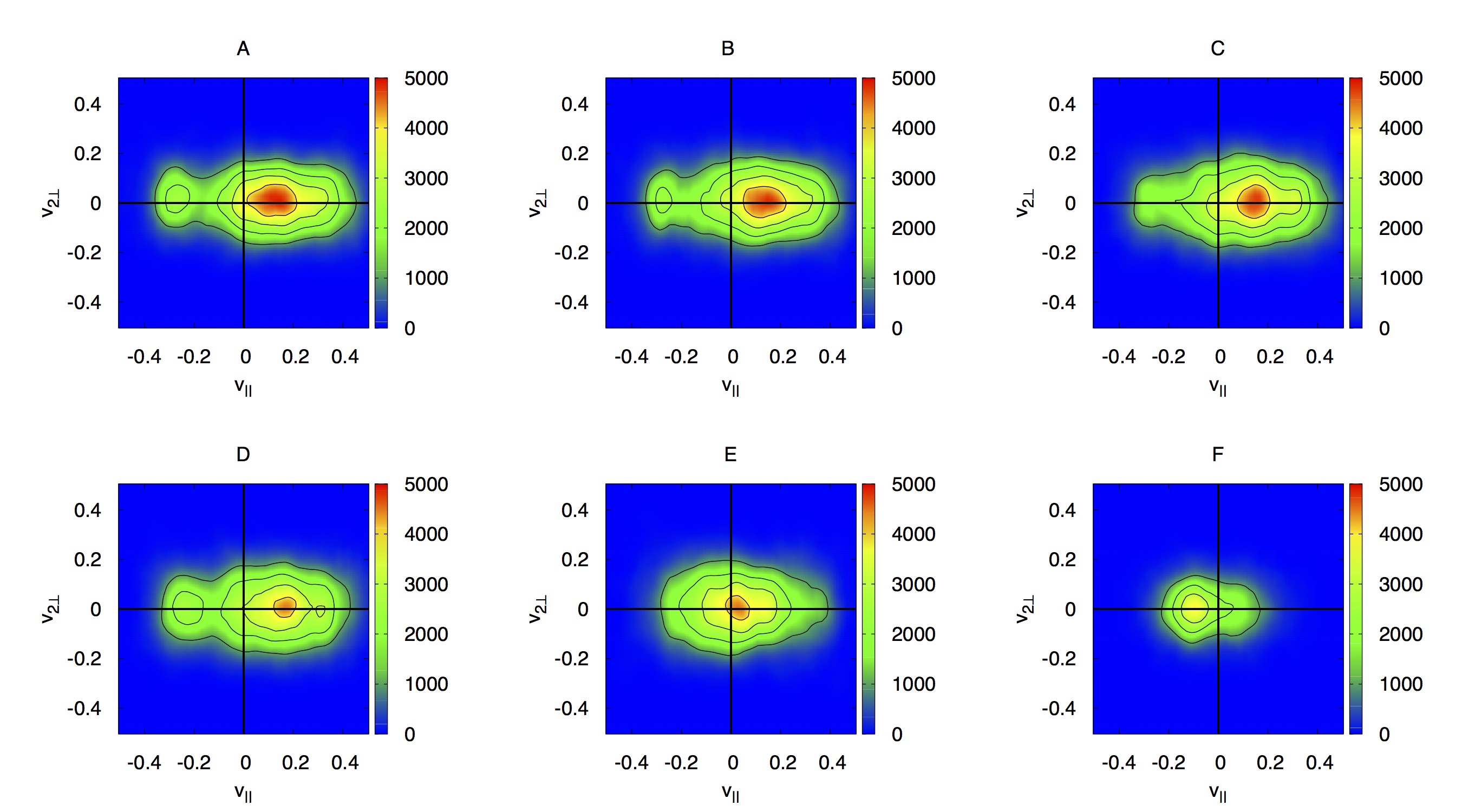}
\caption{Distribution functions for simulation G3 in the $v_{e||}$-$v_{2\perp}$ plane at different locations, see Fig. \ref{Fig:boxes}, where boxes indicate the areas where distribution functions are calculated. See text for further explanation.}
\label{Fig:distribution_functions1}
\end{figure*}
\begin{figure*}
\includegraphics[width=130mm]{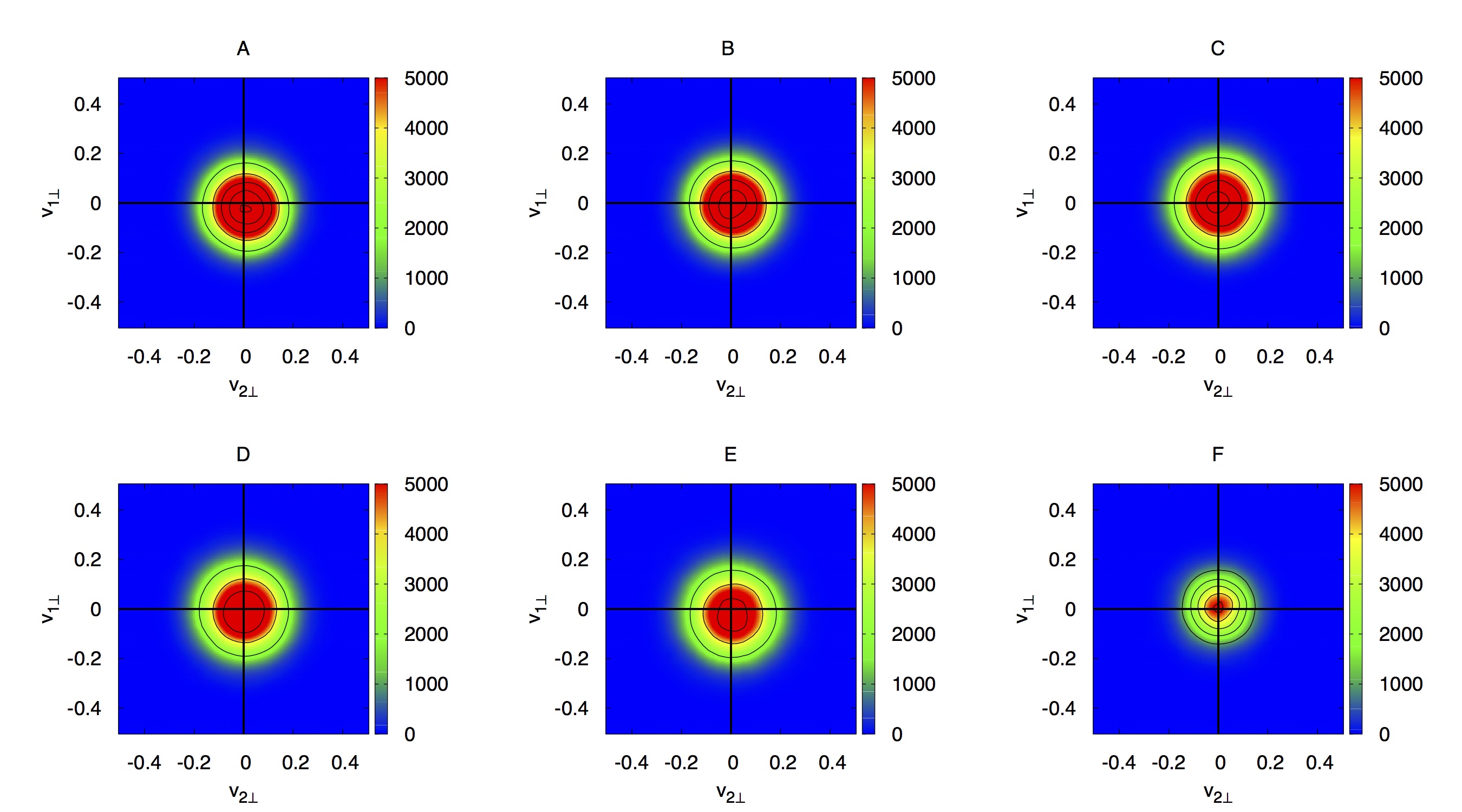}
\caption{Distribution functions for simulation G3 in the $v_{2\perp}$-$v_{1\perp}$ plane at different locations, see Fig. \ref{Fig:boxes} where boxes indicate the areas where distribution functions are calculated.}
\label{Fig:distribution_functions2}
\end{figure*}

The result in \ref{sec:3B} can be explained in terms of the electrostatic field which forms locally due to charge separation. In zero guide field configurations \citet{Chenetal:2015} observed that  an electrostatic field forms close to the reconnection region. They noticed that because the ion gyro-radii are comparable to or larger than the spatial localization width of the electrostatic field ${\bf E}_{es}$, ions can be accelerated or decelerated by ${\bf E}_{es}$, depending on the gyrating ion velocity direction with respect to the electric field itself. The particle acceleration process is not necessarily irreversible so that charge separation can also transfer energy from the plasma to the fields by generating the electrostatic field. This also occurs at the high density separatrices of $B_{0z}=3$, which present strongly negative values of $D_e$, see Fig. \ref{Fig:dissipation1}. Indeed if we follow a typical electron orbit, as shown in Fig. \ref{Fig:orbite00} we see the electrons move from the low density separatrix towards the x-point, where they are strongly accelerated by the parallel reconnection electric field, then they enter the high density separatrix. In Fig. \ref{Fig:boxes} we show a zoom of the left high density separatrix region for the $B_{0z}=3$ case; color coded is the electron density, superimposed are the electron fluid velocity (purple arrows) and the magnetic field lines (black solid). The boxes indicate regions where we calculated the distribution functions. The phase space is projected in the reference frame defined by the vectors $v_{||}= {\bf v} \cdot {\bf B}/| {\bf B }|$, $v_{1\perp}= {\bf v} \cdot {\bf e}_1$ and $v_{2\perp}= {\bf v} \cdot ({\bf e}_1 \times {\bf B}/| {\bf B }| )$, where ${\bf e}_1 = {\bf B}_{P} \times \hat{\bf z}/|{\bf B}_{P}|$, $\hat{\bf z}$ identifies the direction out of the reconnection plane and ${\bf B}_{P}=(B_x,B_y,0)$ is the magnetic field within the plane \cite{Chenetal:2015}. The distribution functions in Fig.\ref{Fig:distribution_functions1} and Fig.\ref{Fig:distribution_functions2} are labelled with capital letters, accordingly to the positions where they are calculated in  Fig.\ref{Fig:boxes}. 
Electron motion results in beamed distribution functions, with high parallel (to the local magnetic field) velocities, as shown in Fig. \ref{Fig:distribution_functions1}.
The distribution function in the plane perpendicular to the magnetic field in Fig.\ref{Fig:distribution_functions2} are isotropic and show a moderate heating in the outflow regions (A-E) with respect to the inflow region (F). Following the electron orbits, moving away from the x-point (from area A to D), we notice in Fig.\ref{Fig:distribution_functions1} the number of electrons populating the beam component (color coded in red) through the parallel acceleration decreases: indeed the electric field decelerates the electrons and accelerate ions in order to restore the charge neutrality. To prove our hypothesis we calculated the work done by the parallel electrostatic field on the electron fluid $W_e=-\int n_e |e| {\bf E} \cdot {\bf v}_e$ where $e$ is the electron charge, and the integral is calculated around the high density separatrix region. We found $W_e$ to be highly negative. The negative beam in the distribution functions it is most probably due to the particles that are reflected from mirror forces.
%
%
\section{Energy partition for different value of the guide field.}\label{sec:III}

\begin{figure}
\includegraphics[width=70mm]{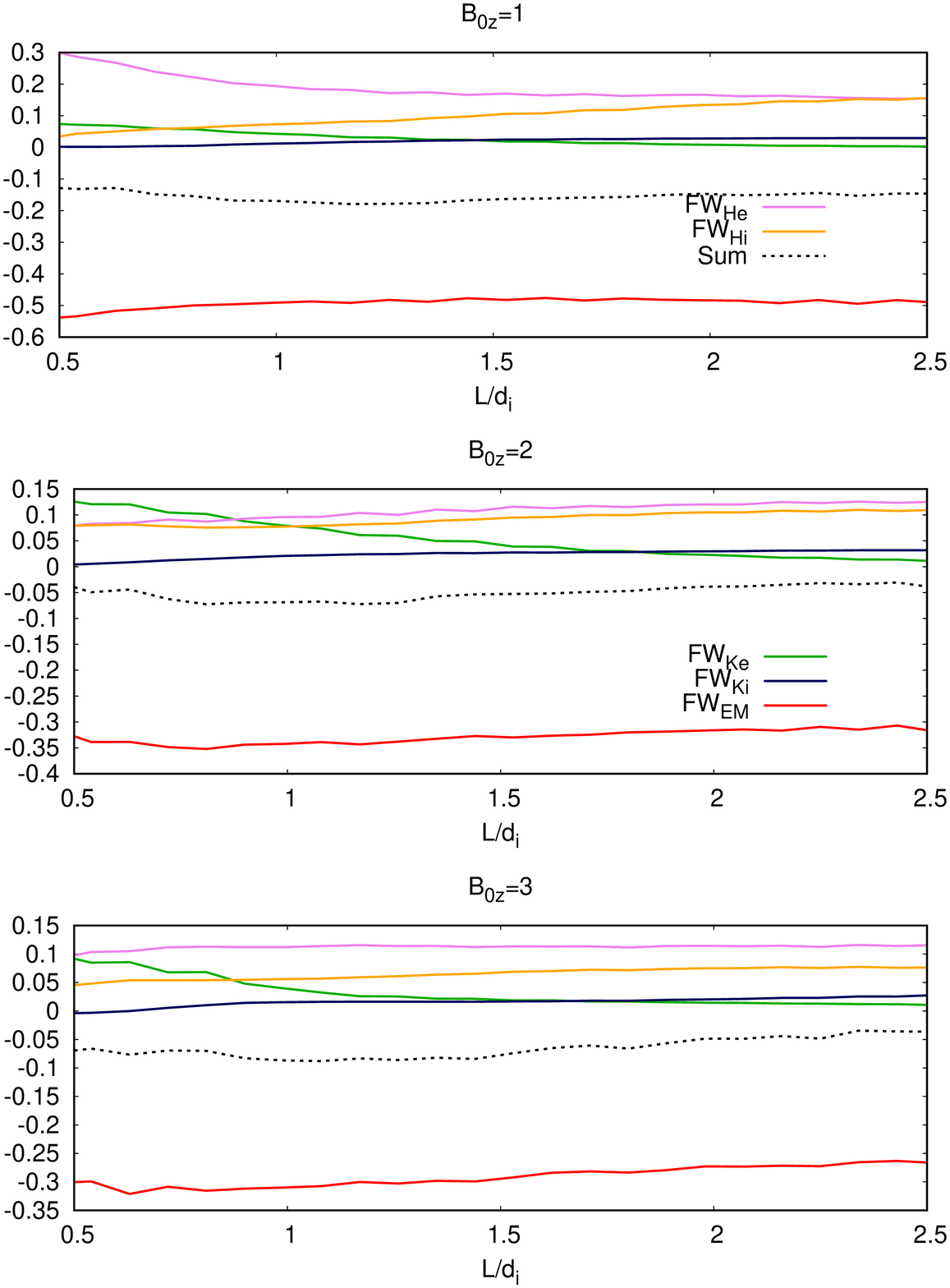}
\caption{Stationary energy balance, Eq. \ref{eq:Energy_balance_fluxes}, integrated over squares centered at the x-point of linear size $L/d_i$, indicated on the horizontal axis. The panels show different guide field configuration (corresponding to simulations $G1-G3$ as indicated in the panels). $FW_{He}$ and $FW_{Hi}$ are the electron and ion internal energy fluxes respectively, $FW_{Ke}$
 and $FW_{Ki}$ are the electron and ion kinetic energy fluxes,
$FW_{EM}$ is the contribution from the Poynting flux ${\bf S}$ and $Sum$ is the sum over all contributions. All the quantities are normalized to the incoming Poynting flux ${\bf S}$ at the upper and lower boundary of the integration area. }
\label{Fig:partition}
\end{figure}

We now proceed to quantify the energy balance described by the energy equation
\begin{eqnarray}
\label{eq:Energy_balance_fluxes}
&&\partial_t \left[\dfrac{E^2+B^2}{8\pi} + \Sigma_s \left(\dfrac{1}{2} n_s m_s {\bf u}^2_s + \dfrac{1}{2} Tr(\bf P)_s\right)\right]\nonumber \\[2ex]
&&+ \nabla \cdot \Sigma_s \left[ \left(\dfrac{1}{2} n_s m_s {\bf u}^2_s+ \dfrac{1}{2} Tr({\bf P}_s) \right) {\bf u}_s\right]\nonumber \\[2ex]
&&+\nabla \cdot \Sigma_s \left[  {\bf P}_s : {\bf u}_s +  c\dfrac{{\bf E}\times{\bf B}}{4\pi}+ {\bf q}_s\right]=0
\end{eqnarray}
where the first three terms are respectively the magnetic, kinetic and internal energy for each species $s$. We considered a pressure tensor ${\bf P}_s$ for each species with $Tr({\bf P}_s) = \Sigma_i P_{ii}$ the trace of the pressure tensor. The term ${\bf S}=c({{\bf E}\times{\bf B}})/{4\pi}$ is the Poynting flux, while  ${\bf q}_s$ are the heat fluxes, whose contribution we neglect. 
We analyze the stationary configuration so that $\partial_t=0$, averaging the fields over a few ion gyro times. Following \citet{YamadaNature2014,Yamada2015} we integrated Eq. \ref{eq:Energy_balance_fluxes} over squares centered at the x-point. In Fig. \ref{Fig:partition} we can see the contribution for the energy fluxes and the Poynting flux ${\bf S}$ for different guide field configuration (corresponding to simulations $G1-G3$ as indicated in the panels). Each line quantifies the flux across the surface (perimeter in a 2D case) of a volume (area) $[L\times L]$, where the linear size $L$, normalized to $d_i$, is indicated on the horizontal axis. $FW_{Hs}= \bigint_V  \nabla \cdot \left[ \dfrac{1}{2} Tr({\bf P}_s) {\bf u}_s + {\bf P}_s \cdot {\bf u}_s\right] d^3 x$    is the internal energy flux for each species $s$, $FW_{Ks}= \bigint_V  \nabla \cdot \left( \dfrac{1}{2} n_s m_s {\bf u}^2_s {\bf u}_s\right) d^3 x$ is  kinetic energy flux, and $FW_{EM}= \bigint_V \nabla \cdot {\bf S} \, d^3 x$.
All the quantities are normalized to the incoming Poynting flux at the upper and lower boundary of the integration area. For the no guide field case please refer to \citet{Yamada2015}. Negative values mean the incoming flux is larger than the outgoing energy flux. As expected the Poynting flux is negative, which means the magnetic energy entering the boxes is converted into other kinds of energies. 
In Fig. \ref{Fig:partition}  we can see for guide field configurations magnetic energy is mainly converted into electron internal energy (pink solid line) and ion internal energy (orange solid line). For $B_{0z}=1$, at scales larger than $2d_i$ the partition between electron and ion internal energy is about the same, while for $B_{0z}=3$  ion heating is about half of electron heating.
This is in qualitative agreement with zero guide field case \cite{Yamada2015}.
The fact that the sum over all the fluxes is not exactly zero, see dashed line in Fig.\ref{Fig:partition}, is due to the fact that we neglected the contribution from the heat fluxes (similar to \citet{Yamada2015}). In addition, even though the configuration is statistically stationary at late times ($t \omega_{ce}>300$),  there remain significant time dependent fluctuations. The latter affects in particular the conversion at very small scales ($L<d_i$) at which  we can see electron kinetic energy gain is comparable with ion internal energy for high guide field configurations. In particular the electron kinetic energy flux can be dominant at small scales ($L<d_i$), e.g. for $B_{0z}=2$, depending on the specific time interval over which we average.
%
%
%


\section{Electron heating for different guide field configurations.}\label{sec:IV} 

\begin{figure*}
\includegraphics[width=170mm]{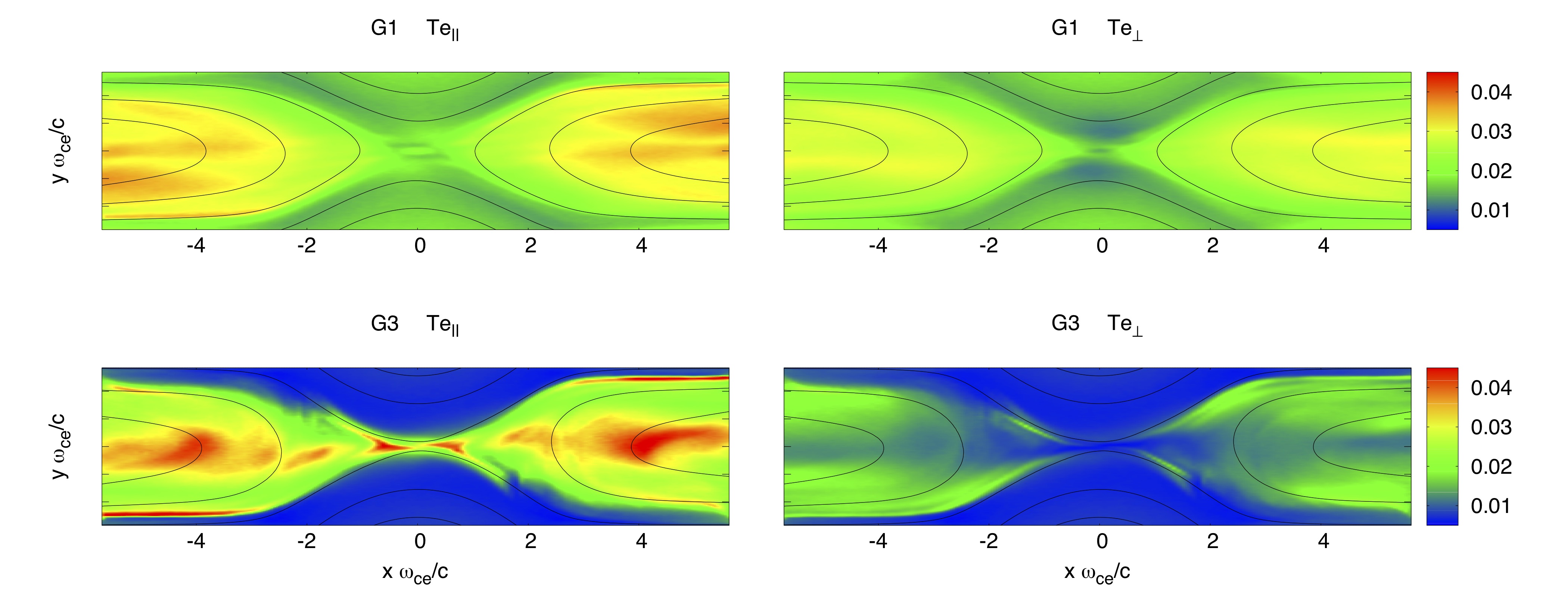}
\caption{2D profiles of parallel and perpendicular temperature, normalized with $m_e c^2$, for simulation G1 (a-b) and G3 (c-d), averaged on $\sim 6$ electron gyro times.}
\label{Fig:temperatures}
\end{figure*}
\begin{figure}
\centering
\includegraphics[width=60mm]{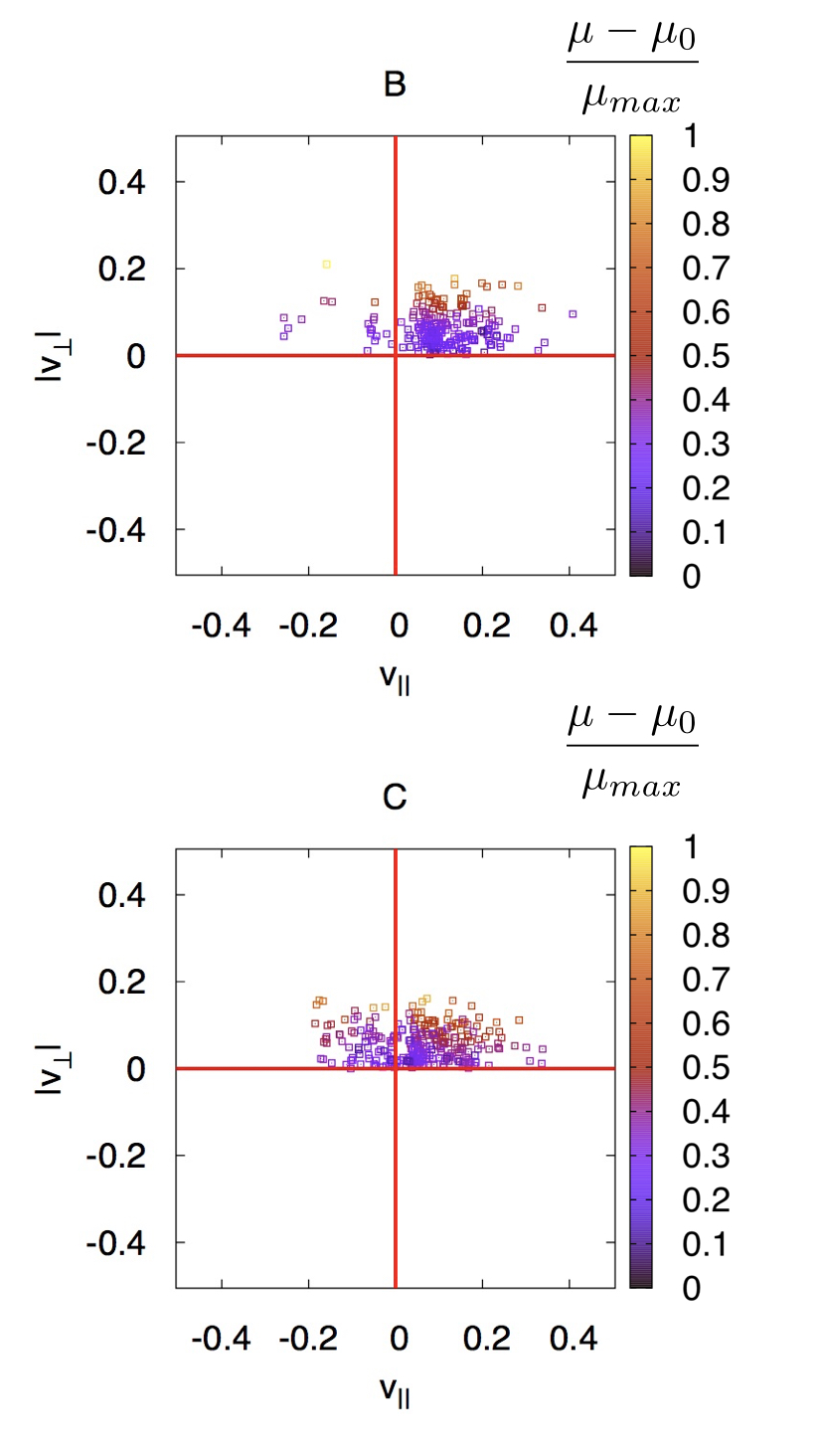}
\caption{Real particle distributions in the phase space plane defined by $v_{||}$ and $|v_{\perp}|$, for simulation G3, i.e. high guide field configuration. Color coded the difference between the magnetic moment $\mu$ at the location where the distribution is calculated and the original location of the tracked particle, normalized to the maximum magnetic moment $\mu_{max}$ in the labelled area.}
\label{Fig:particles}
\end{figure}
\begin{figure}
\includegraphics[width=65mm]{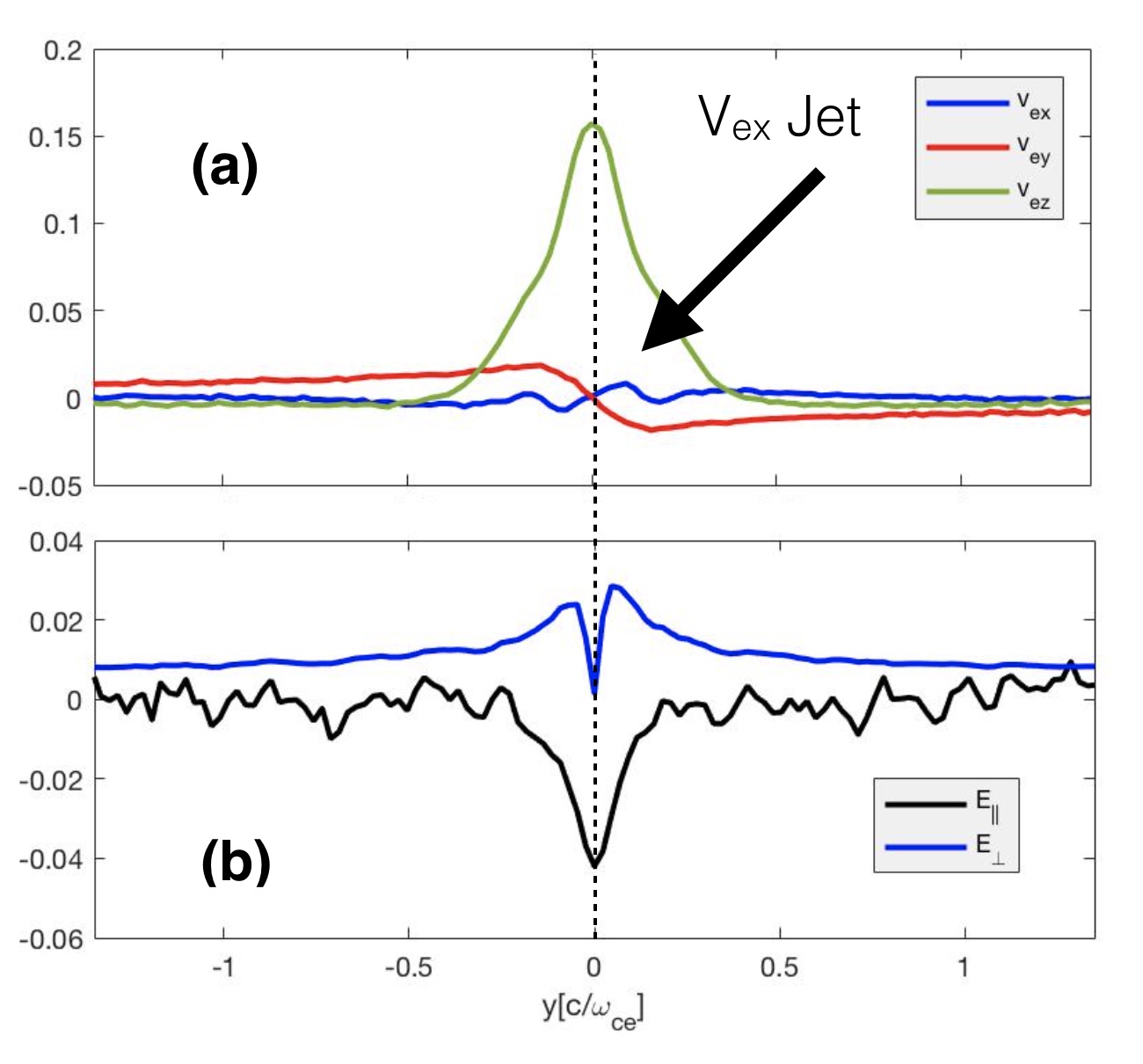}
\caption{1D cut at $x\sim0.4 c/\omega_{ce}$ of the 2D profiles in PASMO code of (a) electron fluid velocity components  (b) parallel and perpendicular electric field components for simulation G1.}
\label{streamings}
\end{figure}

As remarked by \citet{Yangetal2017} the energy transfer ${\bf E \cdot J}$ accounts for both reversible and irreversible energy transfer processes. We now study the electron heating for different guide field configurations. We define the parallel electron temperature as ${T}_{e ||}={\bf P}_e : {\bf B}{\bf B}/n_e$ and the perpendicular electron temperature as ${ T}_{e \perp}={\bf P}_e : ({\bf I}-{\bf B}{\bf B})/2n_e$, so that $T_s$ of the species $s$ is normalized with $m_e c^2$. In Fig. \ref{Fig:temperatures}, the top panels show parallel and perpendicular electron temperature respectively, in the mild guide field configuration, $B_{0z}=1$; the bottom panels show similar 2D profiles for high guide field configuration, $B_{0z}=3$. Depending on the intensity of the guide field, electron heating may occur in the downstream region (low and moderate guide field case) or close to the x-point, along the separatrices (high guide field case). According to panel (a) and (b) we can see that, for low guide field configurations, both parallel and perpendicular temperature rise in a wide downstream region, showing the temperature to be approximately isotropic. Similar patterns can be identified for intermediate guide field configuration $B_{0z}=2$. For $B_{0z}=3$ there is a strong anisotropy, the heated area becomes narrower, closer to the reconnection plane, with very high peaks of parallel temperature. 
This can be explained by the magnetization parameter, defined as $\mathcal{K}=\mathrm{min}(\sqrt{R_B/\rho_e})$, where $R_B$ is the curvature radius and $\rho_e$ is the electron Larmor radius; if  $\mathcal{K}>>1$ particles are fully magnetized \cite{Nortrop1963}. This condition is not satisfied in the downstream region for low guide field configuration: particles can scatter, mixing their pitch angle, so the downstream region becomes isotropic both for parallel as well as for perpendicular temperature. 
Even if the relative difference between peak parallel and perpendicular temperatures is $\sim 50\%$, and the magnetization parameter is expected to be very high, Fig. \ref{Fig:temperatures} (d) shows that perpendicular heating occurs close to the separatrices region. When particles are magnetized the magnetic moment $\mu=m v^2_{\perp}/(2B)$ is most often conserved. From a kinetic analysis \citet{Guo_etal:2017} showed that perpendicular electron heating is mainly due to the non conservation of electron magnetic moment in the separatrix regions.\\
To further understand the behavior of electrons we studied several particles trajectories. Particle trajectories and statistics have been extensively studied both in the antiparallel reconnection (\citet{Egedal_etal:2005}, \citet{Egedal_etal:2008},  \citet{Zenitani_Nagai:2016}) as well as in the mild guide field case (\citet{Pritchett_etal:2006}, \citet{Huang_etal:2010}, \citet{Zenitani_etal:2011}). In our case particle trajectories evolve self-consistently within the plane where magnetic reconnection occurs. As mentioned in the previous section, Fig. \ref{Fig:orbite00} shows two typical electron orbits along magnetic field lines, reaching the reconnection region and then moving away from it along the high density separatrix. We decompose the velocity space in the direction parallel and perpendicular to the magnetic field, defining the vectors $v_{||}= {\bf v} \cdot {\bf B}/| {\bf B }|$ and $v_{\perp}= {\bf v}-v_{||}$. In Fig. \ref{Fig:particles} we show the velocities of the particles within the areas defined in Fig. \ref{Fig:boxes}, as points in the $v_{||}$ and $v_{\perp}$ plane.
%
%
Particles are color coded according to the value of their magnetic moment difference $(\mu-\mu_0)/\mu_{max}$, i.e. the difference from the magnetic moment $\mu_0$ in the upstream region with respect to one in the area we want to analyze $\mu$, normalized to the maximum magnetic moment $\mu_{max}$. From this statistical analysis it appears that, for the particles that populate a high perpendicular velocity area of the distribution function, the magnetic moment is not conserved, so the perpendicular heating is possibly due to unmagnetized particles in this area. Additional explanation to the observed high perpendicular temperature are possible. In particular we would like to suggest a possible fluid explanation.
In Fig. \ref{Fig:boxes} we can see purple arrows superimposed, corresponding to the electron fluid velocity. The presence of ordered sheared electron beams (inflow and outflow) in the high density separatrix region, corresponds to high perpendicular temperature. These sheared flows, not observed in the absence of a guide field, may modify the components of the pressure tensor, as suggested by \citet{DelSarto_Pegoraro:2018}, enhancing the non diagonal terms of the tensor itself. Our conclusion is that the break of the magnetic moment conservation and the shear flows are respectively kinetic and fluid explanation for the electron perpendicular heating observed in the separatrices region. \citet{Wilder_etal:2017}, as discussed in the introduction, investigated a symmetric magnetic reconnection event from MMS with moderate guide field. They observed electron jets in conjunction with a spatially and temporally persistent dissipative parallel electric field. The parallel electric field heats electrons that drift through it. In Fig. \ref{streamings} for simulation G1, we show 1D cuts at $x\sim0.4 c/\omega_{ce}$ of the 2D reconnection plane of the (a) components of the electron fluid velocities and (b) Electric field parallel and perpendicular to magnetic field. Note that the time dependence in reference \citet{Wilder_etal:2017} is a proxy for position of the moving spacecraft, so we can recognize the presence of a similar counter-streaming electron velocity structure in the correspondence of high parallel electric field. As the latter is a signature of guide field component the counterstreaming electron beams are a possible explanation for the jets observed by MMS.
%
%
%
\section{Conclusions}
In this work, we have studied the plasma energization and energy partition in driven two-dimensional symmetric reconnection configurations for different guide field intensities, with emphasis on electron acceleration and heating at the reconnection point and in the close downstream region. We were motivated by laboratory measurements and recent magnetospheric observations, which have shown different features for electron temperature and dynamics from the anti-parallel case. We analyzed the contribution of parallel and perpendicular energy transfers, recovering the same trend and threshold for the transition from the perpendicular dominated energy transfer ($B_{0z}<0.8$) to the parallel dominated one, as found from MRX in agreement with observations by MMS \citet{Fox_et_al:2018}, suggesting a threshold of $B_{0z}=0.6$ Since we were interested in the electron dynamics we studied the energy transfer in the electron fluid frame $D_e$, showing that electron acceleration at the X-point, quantified by the energy deposition, does not continue monotonically moving away, as $D_e$ becomes negative in the separatrix region. Indeed for a high guide field configuration the differential acceleration experienced by electrons and ions along the magnetic field produces charge separation close to the separatrices, as observed by \citet{Guo_etal:2017}, so that an electrostatic field is formed. We calculated the work done by the electric field on the electrons, which is found out to be negative, in order for the plasma to restore charge neutrality. We studied the energy partition between internal and kinetic energies of different species, quantifying the incoming and outgoing energy fluxes in the reconnection region. We found the magnetic energy entering the boxes is converted into plasma energy as expected, mainly into electron internal energy and ion internal energy. Our result is in qualitative agreement with zero guide field case \cite{Yamada2015}, while quantitatively in the guide field configuration more energy goes to electron internal energy with respect to ion energy. \\
Additional features in the high guide field case that are not present in the zero/low guide field case configuration are the well structured electron counter-streaming flows (described in \ref{sec:IIIC}), which we suggest as a possible explanation for the jets observed by MMS \citet{Wilder_etal:2017} These structure are important as a possible mechanism to explain the perpendicular heating observed in high guide field configuration cases.
While in the low guide field case the magnetization parameter in the outflow region is usually smaller than one, explaining electron thermalization in the downflow region, in the high guide field case we would expect the particles to be strongly magnetized.  As shown by the analytical study of \citet{DelSarto_Pegoraro:2018} shear flows may be responsible of changes in the non diagonal terms of the pressure tensor, contributing to the observed perpendicular heating at or near the separatrices. Further investigation of this process will be addressed in a future work. We also confirmed through a statistical study of self consistently evolved particles that the magnetic moment is not conserved for most of the particles populating the high perpendicular velocity region of the distribution function, as first remarked in \citet{Guo_etal:2017}. We conclude that the last two mechanisms can both be responsible for the perpendicular temperature in the separatrix regions seen in high guide field simulations, explaining the heating mechanism respectively in the fluid and kinetic frameworks.\\
Driven stationary reconnection with an open boundary domain appears to be at least qualitatively different from spontaneous periodic setups. For example, in \citet{Drake_et_al:2005}, a double current sheet with guide field $B_{z}/B_0=1$ is simulated. The electron density cavities and temperature patterns resemble our $B_{0z}=3$ case. The quadrupolar structure is also similar to our $B_{0z}=3$ case, suggesting that differences may be due to different boundary conditions or parameters, such for example $v_A/c$, as well as other parameters, suggesting further studies are necessary.
\section{Acknowledgements}
F. Pucci would like to thank Dr. R. Kumar, Dr. D. Del Sarto and Dr. M. Innocenti for interesting discussions on the interpretation of energy partition and electron temperature.
This simulation work was performed by means of the Plasma Simulator at the National Institute for Fusion Science (NIFS) with the support and under the auspices of the NIFS Collaboration Research program (NIFS17KNSS085). This work was supported by: the Strategic International Research Exchange Promotion Program in National Institutes of Natural Sciences, Japan; Grant-in-Aid from the Japan Society for the Promotion of Science (JSPS) Fellows 15J03758, Grant-in-Aid for Scientific Research 15H05750, the Max-Planck Princeton Center for Plasma Physics, funded by the U.S. Department of Energy under Contract No.DE-AC0204CH11466 and NASA under Agreements No.NNH15AB29I and No.NNH14AX631.

\bibliography{Bib}

\end{document}